\documentstyle[12pt,emlines,bezier]{ioplppt}          
\def\tenrm{}
\def\be{\begin{equation}}
\def\ee{\end{equation}}
\def\beq{\begin{eqnarray}}
\def\eeq{\end{eqnarray}}
\def\lb{\label}
\def\nn{\nonumber}

\def\ve{\varepsilon}
\def\dc{\partial}
\def\bc{\begin{center}}
\def\ec{\end{center}}
\def\bt{\begin{tabular}}
\def\et{\end{tabular}}

\def\ba{\begin{array}}
\def\ea{\end{array}}
\def\A0{\stackrel{o}{A}}
\def\det{\mbox{det}}
\def\dim{\mbox{dim}}
\def\sign{\mbox{sign}}
\def\ss{\subsection}
\begin{document}
\jl{1}
\title{On the structure of some typical singularities for
implicit ordinary differential equations}[Singularities
in implicit differential equations]

\author{M V Pomazanov\dag\ftnote{3}{Electronic address
michael@math356.phys.msu.su}}

\address{\dag\
Department of mathematics,
Faculty of Physics,
Moscow State University,
Moscow,
RUSSIA,
119899
}
\begin{abstract}
We study the systems of ordinary differential
equations which are implicit with respect to the higher derivatives, appearing
in the linear form, and their solutions near the singular points. The
invertibility of the higher derivatives reduces to the singular surface
where the theorem of uniqueness and existence is violated.A set of the singular
solution is obtained for the case requiring the minimal number of additional
conditions. The various types of singularities correspond to these cases.
These ``turning'' , ``intersection'' singularities appear in
several physical applications.

\end{abstract}

%
%
\maketitle

\section{Introduction}

The  implicit differential equations of the form
\be\lb{1} A(x,t)\dot x =b(x,t),~~~x\in R^n,~~~t\in R^1 \ee
are considered in this paper
where $A(x,t)$
is a ( n $\times$ n)-matrix and  $b(x,t)$ is a ( n $\times$ 1)-vector function.
The solution of equation (\ref{1}) is called a singular solution if it
includes the points (called singular ones) $x_0,t_0$ where the
main determinant of matrix $A$ vanishes.
Such singular solutions pass through
the  regions where the uniqueness and existence theorem is violated.
We will study behaviour of the solutions in these regions by the asymptotic
methods.  The initial condition for equation (\ref{1}) is set to be
\be\lb{2}
x(t_0)=x_0,~~~\det A(x_0,t_0)=0.
\ee
The surface $\det A(x,t)=0$ could be  called the   singular surface
in the  phase space.

In the special cases the system (\ref{1}) could be a system of Euler-Lagrange
equations defined by the Lagrangian  $L(q,\dot q,t)$,
where $q,\dot q \in R^n$ are vectors of coordinates and  speeds,
respectively.  One can denote the coupling $\{q,\dot q\}^T$ by $x\in R^{2n}$ (a
sign (${}^T$) is a transposition) and present the Euler-Lagrange equations
\[\frac{d}{dt}\frac{\dc L}{\dc \dot q } - \frac{\dc L}{\dc q}=0\]
in the form (1), where the (2n $\times$ 2n)-matrix $A(x,t)$ contains  the
block I (identity (n $\times$ n)-matrix) and the block
\[\frac{\dc^2 L}{\dc \dot q^2}=\{\frac{\dc^2 L}{\dc \dot q_i\dc \dot q_j}\},
i,j=1,2,...n.\]
Therefore, the matrix $A$ has the form
\[A(x,t)=\left(\ba{ll} I & 0\\ 0 & \frac{\dc^2 L}{\dc \dot q^2}\ea\right).\]
The vector function of the right side is defined as
\[b(x,t)=\left(\ba{c} \dot q\\ \frac{\dc L}{\dc q} -
\frac{\dc^2 L}{\dc\dot q\dc t}-\frac{\dc^2 L}{\dc q\dc \dot q}\dot
q\ea\right).\]
In such case the singular surface is determined by the equation
\[H=0,\]
where Hessian $H$ is $H=\det A(x,t)=\det \dc^2/\dc \dot q^2 L.$

The  singular solutions of the Euler-Lagrange equations,
containing the singular points, appear in some physical applications.
For example, as it has been obtained
in the model of post-Gallelelian approximation of
relativistic quasiclassical particle dynamics,
the Lagrangian is singular on some surfaces  of the phase space
\cite{Laser,Pavlov}. Another example of the
singular Lagrangian appears in the model of higher  curvature
string gravity \cite{apom, stas}.

The structure of the paper is the following. In the section 2 the few typical
singularities for the case of degenerated matrix $A$ by the multiplicity 1 are
investigated. Section 3 deals with the simplest case for a degeneration
by two. Prime mathematical demonstration examples are presented in section 4.
In which it is present also the real physical example of typical singularities,
appeared in the black-hole type solutions for the higher curvature string
gravity model.

\section{Several types of the simplest singularities in the
case of the single-valued degeneration}
\ss{Change of coordinates}

Without loss of generality one can consider the system (\ref{1}-\ref{2})
in the following form
\be\lb{3}
\left\{\ba{l} A(x,t)\dot x= B(x,t),\\ x(0)=0,~~~\det A(0,0)=0,\ea\right.
\ee
where $x\in R^n$, $t\in R^1$, $ A(x,t)=\{a_{ij}(x,t)\in C^2[R^n \times R^1]\}$,
$B(x,t)=\{b_i (x,t)\in C^2[R^n \times R^1]\},$  $i,j=\overline{1,n}$.
Let the matrix $\A0=A(0,0)$  be degenerated by the multiplicity 1, i.e.
\be\lb{4}
rang \A0 =n-1.
\ee
The matrix $A(x,t)$ can be rewritten as
\[ A(x,t)=\A0+\sum_{i=1}^{n}x_i A_i(x,t)+tA_t(x,t),\]
where $A_i$, $A_t$ are the ( n $\times$ n)-matrices. Let  some elements of
the matrices $A_i(0,0)$ and $A_t(0,0)$  be nonzero.
We denote by $e_0$ the eigenvector of matrix $\A0$
\[\A0 e_0=0,~~~\|e_0\|=1,\]
and by $e_i$ ($i=\overline{1,n-1}$) the other orthogonal vectors. These vectors
form the basis in
space,  and the following is correct
\[\A0 e_i\neq 0,~~~i=\overline{1,n-1},\]
\[<e_i,e_j>=\delta_{ij},~~~i,j=\overline{0,n-1}\]
(sign $<\cdot,\cdot>$ denotes a scalar product).
The coordinate vector
$x$ can be decomposed in $e_i$  as
\be\lb{5}
x=\tilde y   e_0+\sum_{i=1}^{n-1}z_i e_i.
\ee
In this basis the matrix $\A0$, considered as algebraic operator,
takes the form
\be
\A0=\left(\ba{ll} 0 & \lbrack 0\rbrack^{1}_{n-1} \\
\lbrack 0\rbrack^{n-1}_{1} & \A0_{n-1}\ea\right),
\nn\ee
where $[0]^1_{n-1}$ is the (1 $\times$ n-1)-matrix of the null elements ,
$[0]_{1}^{n-1}$ is the $(n-1\times 1)$-matrix and
(n-1 $\times$ n-1)-matrix $\A0_{n-1}$ is the block of $\A0$
where
\[\det \A0_{n-1}\neq 0.\]
The vector function $B(x,t)$ reduces to a new form, defined
by the basis $e_i$, as
\be\lb{b}
B(x,t)\to (C,d)^T,
\ee
where $C=<e_0,B(x,t)>, d=\{d_i,~i=\overline{1,n-1}\},~~d_i=<e_i,B(x,t)>$.
Taking into
account the doubly continuous differentiability of matrix $A$
elements,
one can present
\[A = \A0+\tilde y  \tilde A +\sum_{i=1}^{n-1}z_i\tilde {A_i}+t\tilde {A_t}+
\lbrack O^2(\tilde y  ,z,t)\rbrack ^n_n,\]
where $\tilde A ,\tilde {A_i} ,\tilde A_t$ are the  constant
( n $\times$ n)-matrices and  the quantity $O^2$
can be understood with preceding notation as
\[O^2(\xi)=\sum_{i=1}^{m}\phi_i(\xi)\xi_i,~~\xi\in R^m,\]
\[\phi_i=O^1(\xi),~~~O^1(\xi)=\sum_{j=1}^{m}\psi_j(\xi)\xi_j,\]
\[\|\psi(0)\|\neq 0,~~~\psi_j(\xi)\in C^2[R^m],~~i,j=\overline{1,m}.\]
                $O^3(\xi),~~O^4,$...
can be defined in the same way .

We denote $D(\tilde y  ,z,t)=\det A(x,t)$
and introduce the new variable $y$ by
\be\lb{y}
y=D(\tilde y  ,z,t).
\ee
\ss{Asymptotic form of equations}

For the first considered case it is possible to assume that
\be\lb{c1}
\frac{\dc D}{\dc \tilde y  }\mid_{\tilde y  ,z,t=0}=r\neq 0.
\ee
Hence
\be\lb{yy}
y=r\tilde y  +<D^0_z,z>+D^0_t t+O^2(\tilde y  ,z,t),
\ee
where $D^0_z=\dc D/\dc z(0,0,0)$ is the constant row vector,
$D^0_t=\dc D/\dc t (0,0,0)$.
In terms of $y$, variable $\tilde y$ takes the form
\be\lb{yty}
\tilde y  =\frac{1}{r}(y-<D^0_z,z>-D^0_t t)+O^2 (y,z,t)
\ee
The matrix $A(x,t)$ can be expressed in terms of $y,z,t$ as
\[ A(x,t)=\left(\ba{ll}{y}/{\det\A0_{n-1}}+O^2(y,z,t) &
\lbrack O^1(y,z,t)\rbrack ^{1}_{n-1} \\
\lbrack O^1(y,z,t)\rbrack ^{n-1}_{1} & A_{n-1}(y,z,t)\ea\right),\]
where $A_{n-1}(0,0,0)=\A0_{n-1},~~~\det\A0_{n-1}\neq 0$.
Denoting by $\bar A$ the matrix of cofactors to the matrix A, one can
write the inverse matrix $A^{-1}$ as
\[A^{-1}=\frac{1}{y}\bar A.\]
Therefore, equations (\ref{3}) can be written as
\be\lb{yr}
y\left({\ba{c}\dot{\tilde y}  \\ \dot z\ea}\right)=\bar A B,
\ee
So, $\bar A$ have the structure
\be\lb{bA}
\bar A(y,z,t)=\left(\ba{ll} \det A_{n-1} & \lbrack O^1(y,z,t)\rbrack ^1_{n-1}\\
\lbrack O^1(y,z,t)\rbrack ^{n-1}_1 & y A^{-1}_{n-1}+\lbrack O^2(y,z,t)\rbrack ^{n-1}_{n-1}\ea\right).
\ee
Taking into account equation (\ref{y}), the derivative of $y$ have
the form
\be\lb{dy}
\dot y   =\frac{\dc D}{\dc\tilde y  }\dot{\tilde y}  +<\frac{\dc D}{\dc z},\dot z>
+\frac{\dc D}{\dc t}.
\ee
Substituting the equations (\ref{yr}) to the expressions (\ref{dy}) and
taking into
account equation (\ref{c1}) and the structure of the  matrix $\bar A$
(\ref{bA}), we obtain
\be\lb{5.1}
y\dot y=r\  \det\A0_{n-1}C^0+\alpha y+<\beta,z>+\gamma t+O^2(y,z,t)
\ee
where $C^0=C(0)$ in (\ref{b}) and $\alpha\in R^1,$ $~\beta\in R^{n-1}$,
$\gamma\in R^1$ are the constants, obtained from the first
derivatives of $\det A_{n-1}(y,z,t),c,d$ at $y,z,t=0$ and from the
vector matrixes $[O^1(y,z,t)]^1_{n-1}, ~~[O^1(y,z,t)]^{n-1}_1 $
disposed in the matrix $\bar A$.
The other part of (\ref{yr}) we rewrite as
\[y\dot z=[O^1(y,z,t)]^{n-1}_1 C^0+y A^{-1}_{n-1}d^0+[O^2(y,z,t)]^{n-1}_{1},\]
and, further, in a form like (\ref{5.1})
\be\lb{6}
y\dot z=y(\Lambda C^0+\A0^{-1}_{n-1} d^0)+(\Delta z+\theta t)C^0+
[O^2(y,z,t)]^{n-1}_1,
\ee
where $d^0=d(0)$ in (\ref{b}) and the coefficients
$\Lambda\in R^{n-1},$ $\Delta \in \{ R^{n-1}\times R^{n-1}\},$
$\theta\in R^{n-1}$ can be calculated using the elements of matrix $\bar A$.

\ss{Turning and intersection}

The equations (\ref{5.1}), (\ref{6}), considered under the initial conditions
$y(0)=0,$ $z(0)=0$, can have two types of singularities. The first type
we define as  ``the turning''.

If $C^0\neq 0$ then equation (\ref{5.1})
can be rewritten in the following simple form
\[y\dot y=F_0+O^1(y,z,t),\]
where $F_0=r\  \det\A0_{n-1}C^0\neq 0$. In such situation equations (\ref{6})
have the form
\[y\dot z=y g^0+[O^1(z,t)+O^2(y,z,t)]^{n-1}_1,\]
where $g^0=\Lambda C^0+\A0^{-1}_{n-1}d^0$ is the constant column vector,
assumed to be nontrivial. That is the reason why
the functions $y(t)$, $z(t)$ have                the unique
asymptotical structure
\[y=\pm\sqrt{2F_0t}+o(\sqrt{F_0t}),~~~z=g^0t+[o(t)]^{n-1}_1\]
defined in the neighbourhood of the null point (Figure 1($a$)).
(The value $o(\cdot )$ may be interpreted in the  generally accepted sense.)

The second type of singularity for the case mentioned above appears under
the condition  $C^0=0$ and is called ``the double intersection''.
The equations (\ref{5.1}), (\ref{6}) have the following general form:
\[\left\{\ba{l} y\dot y=\alpha y+<\beta,z>+\gamma t+O^2(y,z,t),\\
y\dot z=y\A0^{-1}_{n-1}d^0+[O^2(y,z,t)]^{n-1}_1 .\ea\right.\]
Assuming that $y\sim t^\ve+o(t^\ve),$ $z\sim t^\nu+o(t^\nu)$
at $\ve,\nu > 0$, we evidently obtain $\ve=\nu=1$.
     Thus, $y$ and $z$ can be represented as
\[y=\eta t+o(t),~~~z=\xi t+[o(t)]^{n-1}_1\]
where $\eta\in R^1$, $\xi \in R^{n-1}$ are the parameters which must be
calculated. It is obvious that
\[\xi=\A0^{-1}_{n-1}d^0,\]
but for the parameter $\eta$ the square equation is resulted in the form
\[\eta^2=\alpha\eta+G,\]
where $G=<\beta, \A0^{-1}_{n-1}d^0>+\gamma$ is the constant.
Taking into account the sign of discriminant $Dis=\alpha^2-4G$,
we declare the next possible cases:
i) $Dis<0$. Solutions are absent,
ii)$Dis=0$. One solution is present, and
iii) $Dis>0$. Two solutions are present.
The case iii) means that there are two solutions passing through a singular
point and intersecting under the nonzero angle (Figure 1($b$)).

\ss{More complex cases}

Further we will study the case of the singularities when the
condition (\ref{c1}) is broken.
Let us assume that
\be\lb{c2}
\frac{\dc D}{\dc \tilde y  }\mid_0=0.
\ee
We split the coordinate vector $z\in R^{n-1}$ to the new
coordinates  $(\hat y, z)$, where the dimension of new vector $z$ is
decreased by 1 and becomes equal to $\dim\  z =n-2$.
It is assumed for coordinate $\hat y$ that
\[\frac{\dc D}{\dc \hat y}\mid_0=\hat r\neq 0.\]
The part $d$ of the  vector of the right side $B(x,t)$ (\ref{b}) decomposes
to the vector $(\hat c, d)$, where $\dim\  d=n-2$.
By the analogue with (\ref{yy}) we can write now
\be\lb{ynew}
y=\hat r\hat y +<D^0_z,z>+D^0_t t+O^2(\tilde y  ,\hat y,z,t).
\ee
The system (\ref{yr}) has the following form:
\be\lb{sys}
y\left(\ba{c}\dot{\tilde y}  \\ \dot{\hat y} \\\dot z\ea\right)=
\left(\ba{lll}\det A_{n-1} & g_{12} &\lbrack O^1\rbrack^1_{n-2}\\
g_{21} & yl_{22}+O^2(x,t) & yl_{23}+\lbrack O^2\rbrack^1_{n-2} \\
\lbrack O^1\rbrack_1^{n-2} & yl_{32}+\lbrack O^2\rbrack_1^{n-2} &
yl_{33}+\lbrack O^2\rbrack_{n-2}^{n-2}\ea\right)\cdot
\left(\ba{c}c\\ \hat c \\d\ea\right),
\ee
where $g_{21},g_{12}=O^1(x,t), ~~l_{ij}$ are the block components of
the matrix $\bar A$ (\ref{bA}) (for example, $yl_{23}$ is a (1 $\times$ n-2)
row matrix which is a part of block $yA^{-1}_{n-1}$ in $\bar A$ disposed).
The following relation takes place for $\dot y$
\[
\dot y   =\frac{\dc D}{\dc\tilde y  }\dot{\tilde y}  +
\frac{\dc D}{\dc\hat y}\dot{\hat y}+<\frac{\dc D}{\dc z},\dot z>
+\frac{\dc D}{\dc t},
\]
where $\dc D/\dc\tilde y  =O^1(\tilde y  ,\hat y,z,t)$ (see (\ref{c2}))
$\dc D/\dc\hat y=\hat r+O^1(\tilde y  ,\hat y,z,t)$,
$\dc D/\dc z=D^0_z+[O^1(\tilde y  ,\hat y,z,t)]^1_{n-2}$,
$\dc D/\dc t=D^0_t+O^1(\tilde y  ,\hat y,z,t)$.

Expressing the variable $\hat y$ in terms of $y$ by using  (\ref{ynew}) and
substituting the result to the system (\ref{sys}), we can write equations
 (\ref{sys}) in the general form
\be\lb{yr2}
\left\{\ba{l}y\dot{\tilde y}  =C^0 \det\A0_{n-1}+\tilde a \tilde y   +ay+
<b,z>+b_t t+O^2(\tilde y  ,y,z,t),\\
y\dot y=\tilde\alpha \tilde y  +\alpha y+<\beta,z>+\gamma t+O^2(\tilde y  ,y,z,t),\\
y\dot z=C^0(g\tilde y  +[O^1(y,z,t)]^{n-2}_1)+Gy+[O^2(\tilde y  ,y,z,t)]^{n-2}_1,\ea
\right.
\ee
where $\tilde a \in R^1,$ $a\in R^1$, $b\in R^{n-2},$ $b_t\in R^1$,
$\tilde\alpha \in R^1$, $\alpha\in R^1$, $\beta\in R^{n-2}$,
$\gamma\in R^1$, $g\in R^{n-2}$, $G\in R^{n-2}$ are the coefficients
calculated from the matrix $\bar A$.
The initial conditions for the system (\ref{yr2}) are  $\tilde y  (0)=y(0)=0,~$
$ z(0)=0$.
Similarly to the situation considered above, the equations (\ref{yr2})
have a few types of the singular solutions. Here we  investigate some of
the most realizable of them.

\def\ta{\tilde\alpha}
\def\ty{\tilde y}
\ss{Reflecting}

The first type appears under the condition
\[C^0\neq 0,~~~\ta\neq 0,~~~ \|g\|\neq 0.\]
We would like to call this case as ``the reflecting''. We can find the coordinates
$\ty(t),$ $y(t),$ $z(t)$ in the following asymptotical expression
in the same manner as in the above cases
\be\lb{enw}
\ty\sim t^\ve+o(t^\ve), ~~~y\sim t^\nu+o(t^\nu),~~~z\sim t^\omega+o(t^\omega)
\ee
where $\ve>0,\nu>0,\omega>0$ are the unknown coefficients to be
determined. After the substitution of Eq. (\ref{enw}) to (\ref{yr2})
and the equalization of the minimal extents, one obtains the
following system of the  algebraic equations
\be\lb{fc}
\left\{\ba{l}\ve+\nu-1=0,\\
2\nu-1=min(\ve,\nu,\omega,1),\\
\ve+\omega-1=min(\ve,\nu,\omega,1).\ea\right.
\ee
The unique solution for this system is $\ve=1/3,$ $\nu=\omega=2/3$.
Let us introduce the new parameters $\delta\in R^1,$ $\chi\in R^1$ and
$\Delta\in R^{n-2}$  characterizing the singular solutions by
\be\lb{sw}
\ty=\delta\sqrt[3]{t}+o(\sqrt[3]{t}),~y=\chi\sqrt[3]{t^2}+o(\sqrt[3]{t^2}),~
z=\Delta\sqrt[3]{t^2}+[o(\sqrt[3]{t^2})]^{n-2}_1.
\ee
By substituting (\ref{sw}) to (\ref{yr2}) and  setting equal the coefficients
in the  terms with the minimal extents of $t$,  we obtain the equations for
this coefficients
\be\lb{dcD}
\left\{\ba{l}\delta\chi=3 C^0 \det\A0_{n-1},\\
{2}\chi^2=3\ta\delta,\\
{2}\chi\Delta=3C^0 g \delta.\ea\right.
\ee
System (\ref{dcD}) has obviously the unique solution
\[\chi={}^3\sqrt{\frac{9}{2}\ta C^0 \det \A0_{n-1}},~~\delta=\frac{2\chi^2}{3\ta},~~
\Delta=\frac{3C^0\delta}{\chi}g.\]

So, this type means that there is only a single solution passing through a
singular point. The phase point  moves at this solution
to the singular surface $y=0$ and
after the contact with it goes away at the same side (Figure 1($c$)).

\ss{Complicated branching}

The second case defined by (\ref{c2})  appears under the condition
\be\lb{cc2}
C^0\ne 0,~~~~\ta=0,~~~\|g\|\neq 0.
\ee
It is called ``the complicated branching''. For the coefficients
$\ve, \nu, \omega$ we must write the system like (\ref{fc})
\be\lb{sc}
\left\{\ba{l}\ve+\nu-1=0,\\
2\nu-1=min(2\ve,\nu,\omega,1),\\
\ve+\omega-1=min(\ve,\nu,\omega,1).\ea\right.
\ee
We assume here that the  second equation of system (\ref{yr2})
has the form
\[y\dot y =\alpha y +<\beta, z> +\gamma t +\hat\alpha\ty^2+
\ty\cdot O^1 (y,z,t)+O^3 (\ty,y,z,t),\]
where the coefficient $\hat\alpha$, which is assumed nonzero,
can be found from $\bar A$ in (\ref{bA}) like $\alpha, \beta, \gamma$.
This assumption must be added
to (\ref{c2}).
By solving the system (\ref{sc}), we obtain the unique solution
$\ve=1/4,$ $\nu=3/4$, $\omega=1/2$. Therefore, the functions
$\ty(t), y(t)$ and $z(t)$ have the asymptotic form
\be\lb{coef}
\ty=\delta\sqrt[4]{st}+o(\sqrt[4]{st}),~y=\chi\sqrt[4]{st}^3+o(\sqrt[4]{st}^3),
~z=\Delta\sqrt{st}+[o(\sqrt{st})]^{n-2}_1
\ee
where $\delta, \chi,\Delta$ are the unknown coefficients and
sign $s=+1$ or $-1$. This sign indicates  the domain of definition of the
solutions. After substituting
the equations (\ref{coef}) to the system (\ref{yr2}), like in the previous
case (\ref{dcD}),
we  obtain the  system
\be\nn
\left\{\ba{l}\delta\chi=4sC^0 \det\A0_{n-1},\\
{3}\chi^2=4s(\hat\alpha\delta^2+<\beta, \Delta>),\\
\chi\Delta=2sC^0 g .\ea\right.
\ee
The fourth order equation for the coefficient
$\chi$ follows from the above system. It looks like
\be\lb{eqx}
\frac{3}{4}\chi^4-2sC^0<\beta,~~ g>\chi-s\hat\alpha(4C^0 \det\A0_{n-1})^2=0.
\ee
The other coefficients $\delta$ and $\Delta$ can be found
consequently by the formulas
\[\delta=\frac{4sC^0 \det\A0_{n-1}}{\chi},~~~\Delta=\frac{2sC^0}{\chi}g.\]
Now one has the question about the number of real roots of the above
equation.

For answering  this question one can use the well known
Shturm's method \cite{sturm}.
The algebraic form  (\ref{eqx}) is equivalent to
\be\lb{P}
P(\chi)=\chi^4+a_1 \chi+ a_0,
\ee
where the constants $a_1,a_0$ are
\[a_1=-\frac{8}{3}sC^0<\beta,g>,~~~a_0=-\frac{1}{3}s\hat\alpha
(8C^0 \det \A0_{n-1})^2.\]
In order to find the number of real roots,
we must  construct the Shturm's
sequence and then count up the difference between the number
of the alternations of the signs in this sequence at $x=\pm\infty$.
Using the Shturm's algorithm, we obtain that the number of roots of
the equation (\ref{P}) depends on sign of the expression
\[R=a_1^4-\frac{256}{27}a_0^3.\]
When $R<0$, roots do not exist. When $R=0$ only one root (\ref{P})
can exist and when $R>0$ two roots can be found. In our case $a_1=s\hat a_1$,
$a_0=s\hat a_0$ and sign $s$ can be positive or negative.

Therefore, we can declare the next possibilities:
\be\lb{stas}a)~a_1^4<\frac{256}{25}\mid a_0\mid^3,
~~~b)~a_1^4>\frac{256}{25}\mid a_0\mid^3,
~~~c)~a_1^4=\frac{256}{25}\mid a_0\mid^3,\ee
for which the various types of the singular solutions of the differential equations
(\ref{yr2}) can result. The existence of the real singular solutions
for the item (\ref{stas}a) determines the sign $s$ as
\[s=\sign (\hat \alpha).\]
In this case the singular solution of equations (\ref{yr2}) has the
``turning''-like type and can exist only at $t>0$ or $t<0$ (Figure 1($d_1$)).

In the item (\ref{stas}b) the sign
$s$ could be chosen positive or negative.
The  solution like the ``turning''
exists for each of the possible signs. Each solution
contains two branches defined by two roots of the equation (\ref{eqx}).
This singular solution,
placed on both sides of the point $t=0$ (Figure 1 ($d_2$)).

The item (\ref{stas}c) differs from
(\ref{stas}b) with R is equal to zero  at
$\sign ( t)=s=-\sign (\hat\alpha)$. There is only one
branch defined by a single root of equation (\ref{eqx}) for this $\sign ~t$.
 There are two roots (\ref{eqx}) in the  other side
of $t=0$ at $\sign ( t)=+\sign (\hat\alpha)$,
so, there are two branches of the singular solution.
So,  in this case three branches pass through the singular point.

\ss{Triple intersection}

The third case defined by (\ref{c2}) appears under
additional condition
\[C^0=0.\]
It can be  called ``the triple intersection''. The differential
equations (\ref{yr2}) are now  rewritten in the form
\be\lb{yy3}
\left\{\ba{l}y\dot{\tilde y}  =\tilde a \tilde y   +ay+
<b,z>+b_t t+O^2(\tilde y  ,y,z,t),\\
y\dot y=\tilde\alpha \tilde y  +\alpha y+<\beta,z>+\gamma t+O^2(\tilde y  ,y,z,t),\\
y\dot z=Gy+[O^2(\tilde y  ,y,z,t)]^{n-2}_1.\ea
\right.
\ee
It is easy to establish that
$\ve=\nu=\omega=1$ using the analysis of extents.
Therefore, we have the right to suggest the
following behaviour
\be\lb{coef2}\nn
\ty=\delta t+o(t),~y=\chi t+o(t),
~z=\Delta t+[o(t)]^{n-1}_1,
\ee
where obviously
\[\Delta=G.\]
The unknown coefficients $\delta, \chi$ can be obtained from the
algebraic equations
\be\lb{third}
\left\{\ba{l}\chi\delta=\tilde a\delta+a\chi+F,\\
\chi^2=\ta\delta+\alpha\chi+f,\ea\right.
\ee
where $F=<b,G>+b_t,~~~f=<\beta,G>+\gamma$.
We do not pay attention to any degenerate case
of solution (\ref{third}).
In general, the system (\ref{third}) is equivalent the algebraic third
order equation. Hence, it can have three (Figure 1(e)), two or one root.

\section{The double-valued degeneration}
\ss{Equations in neighborhood of a singular point}

The double-valued degeneration means that
\be\lb{rang2}
rang \A0=n-2.
\ee
Most generally in this case, there are two independent egenvectors $e_1, e_2$
corresponding to the zero
egenvalue of the matrix $\A0$. Let us denote $e_i (i=\overline{3,n})$ other basis vectors
orthogonal to $e_1, e_2$.
As in  (\ref{5})
\[x=e_1 y_1+ e_2 y_2+\sum_{i=3}^n e_i z_i.\]
The (n $\times$ n) matrix $\A0$ has the representation
\be
\A0=\left(\ba{ll} \lbrack 0 \rbrack^2_2 & \lbrack 0\rbrack ^{2}_{n-2} \\
\lbrack 0\rbrack^{n-2}_{2} & \A0_{n-2}\ea\right),
\nn\ee
where $\det \A0_{n-2}\neq 0$. So, the matrix $A(x,t)$ has the block form
\def\l{\lambda}
\be\lb{A2}
A(x,t)=\left(\ba{ll} {\ba{ll}\l_{11}&\l_{12}\\ \l_{21}&\l_{22}\ea} &
\lbrack O^1(y_1,y_2,z,t)\rbrack ^{2}_{n-2} \\
\lbrack O^1(y_1,y_2,z,t)\rbrack ^{n-2}_{2} & \A0_{n-2}+
\lbrack O^1(y_1,y_2,z,t)\rbrack ^{n-2}_{n-2}\ea\right),
\ee
where $\l_{ij}=O^1(y_1,y_2,z,t)$.  The vector of right part $B(x,t)$
reduces to the form
$(C_1(x,t),~C_2 (x,t), d(x,t))^T$, where
$C_i,d_i=<e_i, B>,~~i=\overline{1,n}$. The main determinant
$\det A(x,t)$ is
\[\det A(x,t)=\l\cdot\\det\A0_{n-2}+O^3(x,t),\]
where $\l=\l_{11}\l_{22}-\l_{21}\l_{12}$.
Let us choose the orthogonal egenvectors $e_1,e_2$ in order to present
the following form for the value $\l$
\be\lb{lambda}
\l=\l_1 y_1^2+\l_2 y_2^2+\sum_{i=1}^{2}y_i O^1 (z,t)+O^2(z,t)+
O^3(y_1,y_2,z,t)
\ee
(it always could be performed by the rotation of $e_1,e_2$).
The matrix of the cofactors to $A$ is
\be\lb{bA2}\fl
\bar A(x,t)=\left(\ba{ll} {\left(\ba{ll}\l_{22}&-\l_{12}\\ -\l_{21}&\l_{11}\ea
\right)}\cdot \det\A0_{n-2}+\lbrack O^2\rbrack ^{2}_{2} &
\lbrack O^2(y_1,y_2,z,t)\rbrack ^{2}_{n-2} \\
\lbrack O^2(y_1,y_2,z,t)\rbrack ^{n-2}_{2} & \l\A0^{-1}_{n-2}+
\lbrack O^3\rbrack ^{n-2}_{n-2}\ea\right),
\ee
and the differential equations (\ref{yr}) are rewritten as
\[\l\left(\ba{c}\dot y_1\\ \dot y_2\\ \dot z\ea\right)=\bar A
\left(\ba{c}C_1(y_1,y_2,z,t)\\ C_2(y_1,y_2,z,t)\\ d(y_1,y_2,z,t)\ea\right).\]
So,
\be\lb{y12}
\left\{\ba{ll}\l\dot y_1 &=(\l_{22}C^0_1-\l_{12}C^0_2)\det \A0_{n-2}+O^2(y_1,y_2,z,t),\\
\l\dot y_2 &=(-\l_{21}C^0_1+\l_{11}C^0_2)\det\A0_{n-1}+O^2(y_1,y_2,z,t),\\
\l\dot z &=[O^2(y_1,y_2,z,t)]^{n-2}_{2}\cdot{\left(\ba{c}C_1^0\\ C_2^0\ea\right)}+
\l\A0^{-1}_{n-2}d^0\\ &+[O^3(y_1,y_2,z,t)]^{n-2}_1,\ea\right.
\ee
where $C^0_1=C_1(0,...),C^0_2=C_2(0,...),d^0=d(0,...)$.

\ss{The simplest case}

We'll discuss here only one case appearing at
\be\lb{c1c2}
\mid C^0_1\mid+\mid C^0_2\mid\ne 0,
\ee
and call it ``the linking of turnings''.

It is easy to see that the variables $y_1,y_2,z$ have the following unique
structure
\[y_i\sim  t^{1/2},~~~z\sim t.\]
By defining  the system (\ref{y12}) more concretely, we write it as
\be\lb{y12s}
\left\{\ba{ll}\l\dot y_1 &=a_1 y_1+b_1 y_2+ O^1(z,t)+O^2(y_1,y_2,z,t),\\
\l\dot y_2 &=a_2 y_1+b_2 y_2+ O^1(z,t)+O^2(y_1,y_2,z,t),\\
\l\dot z &=g_{11}y_1^2+g_{12}y_1 y_2+g_{22}y_2^2+\l\A0^{-1}_{n-2}d^0\\
&+\sum_{i=1}^{2}y_i [O^ 1(z,t)]^{n-2}_1
+[O^2(z,t)]^{n-2}_1+[O^3(y_1,y_2,z,t)]^{n-2}_1,\ea\right.
\ee
where $a_i,b_i\in R^1,g_{ij}\in R^{n-2}$ are the coefficients which can
be obtained from actual form of $\l_{ij}$, from the $O^2$ --- blocks of
the matrix $\bar A$ (\ref{bA2}) and from $C^0_i,~~~i,j=1,2$.
Thus the following expressions take place:
\be\lb{et12}
y_1=\eta_1\sqrt{st}+o(\sqrt{st}),~~~y_2=\eta_2\sqrt{st}+o(\sqrt{st}),
~~~z=\Delta t+[o(t)]^{n-2}_1
\ee
where the coefficients $\eta_i\in R^1,\Delta\in R^{n-2}$ must be calculated.
The coefficient $s$ is the sign and can be equal only +1 or -1 like
the case (\ref{c2}). By substituting (\ref{et12}) to
(\ref{y12s}) and setting equal the coefficients at the terms with minimal
extents of variable $t$, we obtain the algebraic system \be\lb{e12e}
\left\{\ba{l}(\l_1\eta_1^2+\l_2\eta_2^2)\eta_1=2s(a_1\eta_1+
b_1\eta_2),\\
(\l_1\eta_1^2+\l_2\eta_2^2)\eta_2=2s(a_2\eta_1+
b_2\eta_2).\ea\right.
\ee
If the values of $\eta_1,\eta_2$ was known, the
coefficients $\Delta$ can be determined by the formula
\be\lb{Delta}
\Delta=\A0^{-1}_{n-2}d^0+
\frac{(g_{11}\eta_1^2+g_{12}\eta_1\eta_2+g_{22}\eta^2_2)}
{(\l_1\eta_1^2+\l_2\eta_2^2)}
\ee
followed from the third equation (\ref{y12s}) and (\ref{lambda}).
Let  assume that $b_1$ is not equal to zero (or $a_2\neq 0$).
 The relation between
$\eta_1$ and $\eta_2$ has a form of square equation, as  follows
from equations (\ref{e12e}):
\be\lb{sq}
b_1\eta^2_2+(a_1-b_2)\eta_1\eta_2-a_2\eta_1^2=0.
\ee
It  means that real roots of equation (\ref{sq}) can exist
under the condition $Dis\ge 0$, where
$Dis=(a_1-b_2)^2+4b_1a_2$
is the discriminant of  equation (\ref{sq}). Thus the relation is
\[\eta_2=g^\pm\eta_1,\]
where $g^\pm=(-a_1+b_2\pm\sqrt{Dis})/(2b_1)$ and the coefficient
$\eta_1$ can be determined by
\[\eta_1^2\cdot(\l_1+\l_2(g^\pm)^2)=s(a_1+b_2\pm\sqrt{Dis}).\]
Let us satisfy the condition
\[\l_1+\l_2(g^\pm)^2\neq 0\]
which must be added to the condition (\ref{c1c2}) with
the restriction $\mid b_1\mid +\mid a_2\mid \neq 0$.

In this way, sign $s$ must be chosen as
\[s=\sign(R^\pm),\]
where $R^\pm=(a_1+b_2\pm\sqrt{Dis})/(\l_1+\l_2(g^\pm)^2)$ and by joining
of them, the singular solution has the next kind
\[\left(\ba{c}y_1(t)\\ y_2(t)\ea\right)=\pm
\left(\ba{c}\sqrt{R^\pm t}\\ g^\pm \sqrt{R^\pm t}\ea\right)+
[o(\sqrt{R^\pm t})]^2_1,\]
\[z=\Biggl(\A0^{-1}_{n-2}d^0+({\l_1+\l_2(g^\pm)^2})^{-1}\cdot
({g_{11}+g^\pm g_{12}+(g^\pm)^2 g_{22}})\Biggr)t+[o(t)]^{n-2}_1 .\]

The singular solution passing through the singular point is a linking of two
turnings at $Dis>0$. Moreover, if the signs of the value $R^\pm$ are identical,
the both turnings are
placed on the same side of point $t=0$ (Figure 1($f_1$)). Else they are placed
on the various sides (Figure 1($f_2$)).
Only one turning exists at $Dis=0$.

We have considered here  only the typical case taking place under the
condition (\ref{rang2}), but there are  a few degenerate cases,
for example, one of them can appear at $R^{+}=0$ or $R^{-}=0$.
We have restricted oneself only to the general situation.

\section{Examples and discussion}
\ss{A few prime examples}


By using the prime model examples we would like to
demonstrate several types of
the singularities discussed above. For singular differential equation
\be\lb{ex1}
x\dot x =t
\ee
considered at the initial condition placed on the singular
surface $x=0$
\[x(t_0)=0,\]
we easily determine that under the
condition $t_0\neq 0$  ``the turning''
takes place:
\[x(t)=\pm\sqrt{2(t-t_0)t_0}+o(\sqrt{(t-t_0)t_0}).\]
At $t_0=0$ the ``intersection''
takes place
$x=\eta t+o(t)$, where $\eta=\pm 1$. It is easy to solve understand
Eq. (\ref{ex1}) directly. This solution is $x^2=t^2+t_0^2.$

For the equation
\[x\dot x=-t,\]
under the initial condition mentioned above, we have ``the turning'' at $t_0\neq 0$.
However, at $t_0=0$ we do not obtain ``the intersections'' case,
because the real roots of square equation for the parameter $\eta$
do not exist.
It is easy to see from the direct solution  $x^2+t^2=t^2_0,$
that it does not pass through the singular point $x(0)=0$.

To demonstrate so-called  ``the reflection'' case, we'll turn into the next primary
example of system of two differential equations
\be\lb{ex2}
\left\{\ba{c}y\dot x = -1,\\
(y-x)\dot x+\dot y =0,\ea\right.
\ee
with the initial conditions $x(0)=y(0)=0$ placed on the singular
surface $ D=\det ~A=y=0.$
The matrix $\A0$ of system (\ref{ex2}) is
\[\A0=\left(\ba{ll} 0 & 0 \\ 0 & 1\ea\right)\]
and egenvector is $e_0=(1, 0)^T$, $ \A0 e_0=0.$ The vector function of
the right part is $B=(-1, 0)^T$. The vector $\nabla D$ is
$\nabla D=(\dc D/\dc x,\dc D/\dc y )=(0,1)$, so,
$r=(e_0, \nabla D)=0$, but $<e_0,b>\neq 0$.
Thus, the next structure of the singular solution
 may take place in the most common case
\be\lb{xy2}\nn
x=a\sqrt[3]{t}+o(\sqrt[3]{t}),~~y=b\sqrt[3]{t^2}+o(\sqrt[3]{t^2}).
\ee
By substituting this to the equations (\ref{ex2}), we  can obtain
the following algebraic equations for the coefficients $ a,b $
\[\frac{1}{3}ab=-1, ~~~a^2=2b,\]
which have the unique roots $a=-\sqrt[3]{6},~~b=\sqrt[3]{9/2}.$
The direct solution of the system (\ref{ex2}) gives the functions
$x(t),~y(t)$ in the implicit form
\be\lb{sol2}
t=e^{-x}-(1-x+\frac{x^2}{2}),~~y=t+\frac{x^2}{2}.
\ee
The first equation  (\ref{sol2}) is rewritten as
\[t=\frac{x^3}{3!}(-1+\frac{3!x}{4!}-\frac{3!x^2}{5!}+...),\]
from which it is easy to see the unique structure of function $x(t)$,
and further $y(t)$, near the singular point $x(0)=y(0)=0$.
It coincides entirely to (\ref{xy2}).

\ss{Black Hole example and discussion}

The singular solutions of system (\ref{1}) can be very diverse.
 We considered just the cases which have required  the  least number
of additional conditions.
In our opinion, the smaller is the number of additional conditions, the more
probably the case may occurs in the nature.

In the model of higher  curvature
string gravity the four-dimensional black hole
solutions are generated by the low energy string effective action:

\be\lb{act}
S=\int\d^4 x L=\frac{1}{16\pi}\int\d^4 x\sqrt{-g}~[m^2_{\rm Pl}(-R+2\dc_\mu\phi
\dc^\mu\phi)+\lambda e^{-2\phi} S_{\rm GB}],
\ee
where $R$ is the  scalar curvature; $\phi$ is the dilaton field; $m_{\rm Pl}$
is the Plank mass; $\lambda$ is the string coupling parameter.
The latter describes
Gauss-Bonnet (GB) contribution  to the action (\ref{act}). For consideration
of the static,
asymptotically flat, spherically symmetrical black-hole-like solutions,
the  most convenient choice of metric is
\be\lb{met}
\d s^2=\Delta \d t^2-\frac{\sigma^2}{\Delta}\d r^2-f^2(\d\theta^2+
\sin^2\theta\d\varphi^2),
\ee
where functions $\Delta,~\sigma$, $f$ and the
dilaton function $\phi$ depend only on the  radial
coordinate r. In this metric the Lagrangian, transformed to the
most convenient for analysis form, is
\be\lb{lag}
L(\Delta',f',\phi',\Delta,\phi,\sigma,f)=L_0+\lambda L_{\rm GB},
\ee
where $L_0=m_{\rm Pl}^2[\Delta'f'f+\Delta(f')^2+\sigma^2-
\Delta f^2 (\phi')^2]/\sigma,$ $L_{\rm GB}=4 e^{-2\phi}\phi'
[\Delta\Delta'(f')^2-\Delta'\sigma^2]/\sigma^3$
and the stroke denotes $\d/\d r$.

The main determinant of Euler-Lagrange equations followed from (\ref{lag})
in the curvature gauge $f(r)=r$ has the structure
\[D_{main}=\Delta(A\lambda^2\Delta^2+B\lambda\Delta+C)~~\mbox{where}\]
\[A=(-32)e^{-4\phi}\sigma^2[(4\phi'^2r^2-1)\sigma^2 m_{\rm Pl}^2
+\lambda\cdot 12 e^{-2\phi}\Delta'\phi']\]
\[
B=(-32)e^{2\phi}\sigma^4 [\sigma^2\phi'm_{\rm Pl}^4 r^3+\lambda 2
e^{-2\phi}\sigma^2 m_{\rm Pl}^2-\lambda^2\cdot 8 e^{-6\phi}\Delta'\phi']
\]
\[ C=2\sigma^6 [-\sigma^2 m_{\rm Pl}^6 r^4+\lambda^2\cdot 32 e^{-4\phi}
m_{\rm Pl}^2(\sigma^2+2\Delta'r)+\lambda^3 64 e^{-6\phi}\Delta'\phi']\]

At $\lambda=0$ (GB term is absent) the asymptotically flat solution
followed from the Euler-Lagrange ( Einstein ) equations is well known
Schwarzschild's one. This solution is
\[
\Delta=1-\frac{2M}{r},~~\sigma=1,~~\phi={\rm const}.
\]
The domain of its definition is $r=(0,+\infty)$.
This solution has ``the intersection'' type singular point $r_{h}$
at the event
horizon $\Delta=0$. Here the Shcwarzschild's solution intersect with
symmetrical solution which usually has been rejected because it hasn't
physical sense (Figure 2 curve (c)).  But in a case
of nonzero, positive $\lambda$, under the regular event horizon
$(\Delta(r_{\rm h})=0)$, the Lagrangian (\ref{lag}) has the further
singular point at $r=r_{\rm s}<r_{\rm h}.$
This singular ``turning'' point forms by passing trajectory of solution from
the singular surface $A\lambda^2\Delta^2+b\lambda\Delta+C=0$, which is
nontrivial only at $\lambda\ne 0$.
The domain of definition of the new solution
becomes $(r_{\rm s},+\infty)$ and the solution is subjected to a turning showed
in Figure 2. This singularity is absent in the first order curvature gravity
and it is found  as a surprise for the classical gravity.

The investigation, performed in the present work, may be efficient for analysis
of the compound system of the ordinary differential equation with
the implicit linear higher  derivatives when it is hard to obtain
the direct solution. The investigation of the  equations (\ref{1})
by the numerical methods can be made without the significant difficulties
only on the range of invertibility of matrix $A$. In order to extend
the solution through the singular point, it is necessary to know
the structure of the singularity. In the Appendix to Ref. \cite{apom}
 the numerical method of integration the equations (\ref{1}) by additional
parameter is discussed. It allows passing through a singular
``turning'' point automatically.

\ack
The author would like to thank S.Alexeyev for
useful discussions on the subject of this work.

\newpage
\pagestyle{empty}
\Figures
\Figure{
The typical singularities in implicit differential equations.
}

\Figure{Metric functions $\Delta$ and $\sigma$ versus the radial
coordinate $r$ when the event horizon value $r_{\rm h}$ is equal to
20.0 Plank unit values (P.u.v.). The curve (a) is calculated with
$\lambda=1$, the curve (b) is calculated with $\lambda=0$
(Schwarzschild's solution).
The curve (c) is  the nonphysical branch
passing from point $r_{\rm h}$.
The curve (d)
shows $\sigma(r)$ function at $\lambda=1$.
The arrows pointed $r_{\rm s}$ shows the
positions of $r_{\rm s}$-singularity.}

\newpage
\special{em:linewidth 0.4pt}
\unitlength 1.00mm
\linethickness{0.4pt}


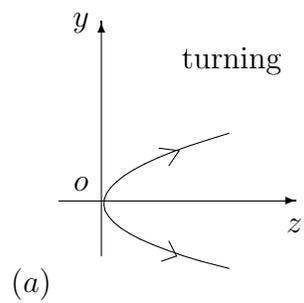
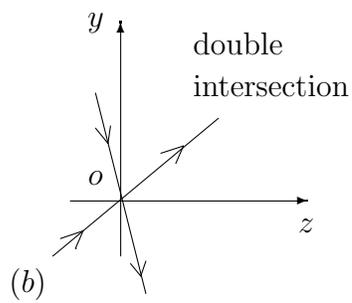
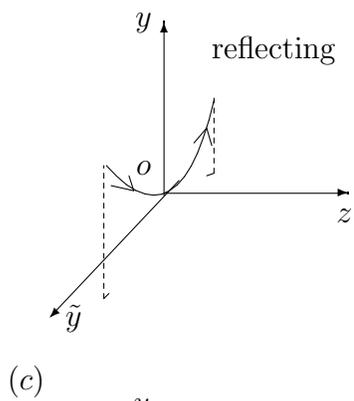
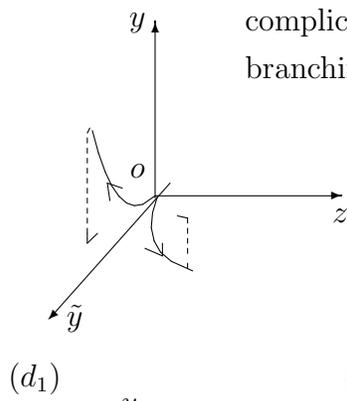
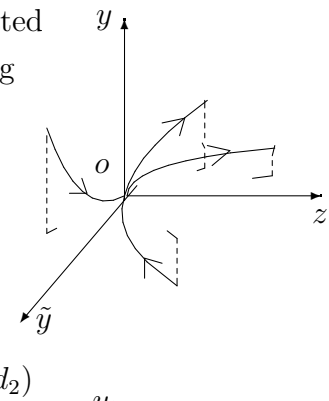
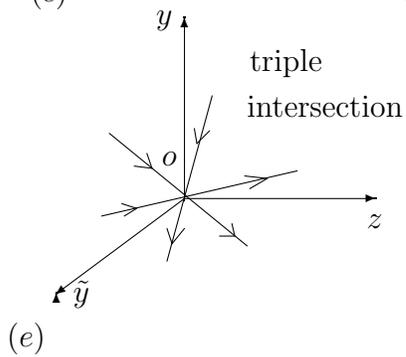
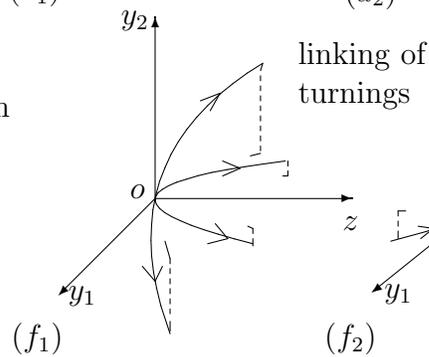
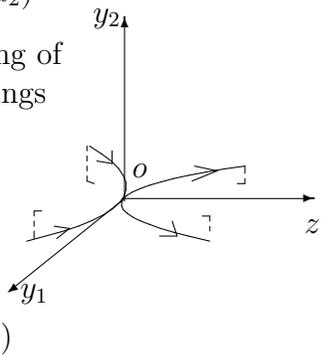
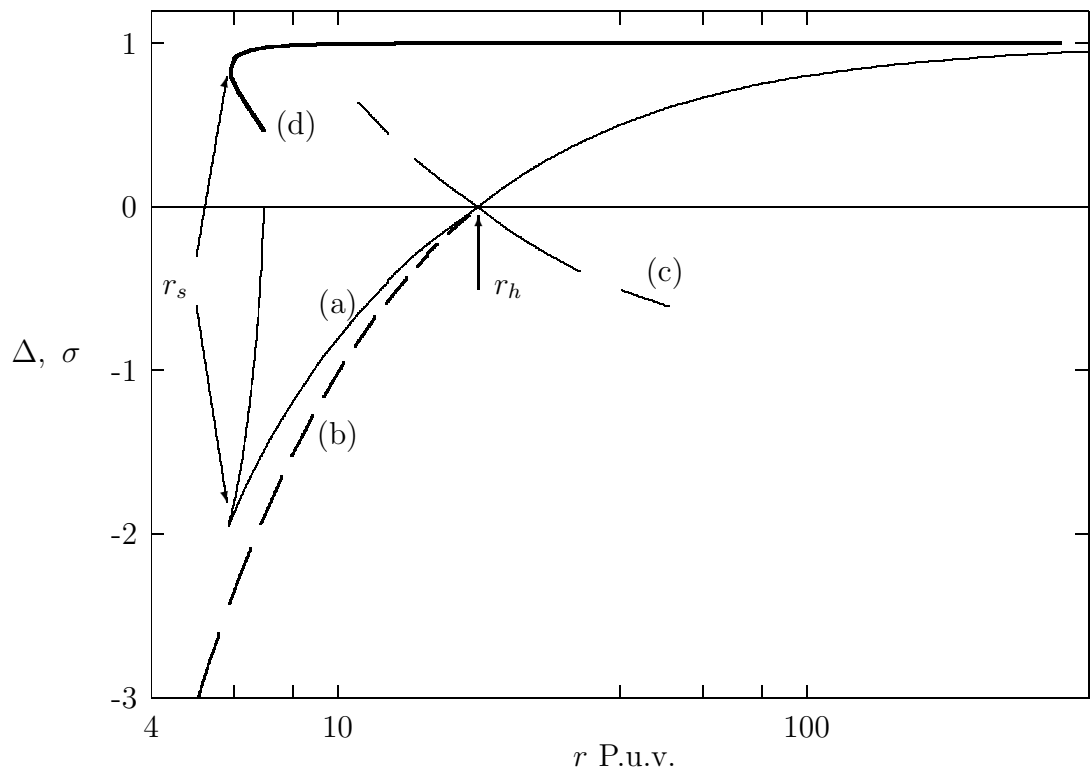
\end{document}